# Validation of a smartphone app to map social networks of proximity

Tjeerd W. Boonstra[a,b,*], Mark E. Larsen[a], Samuel Townsend[a], and Helen Christensen[a]

[a] Black Dog Institute, University of New South Wales, Sydney, Australia
[b] QIMR Berghofer Medial Research Institute, Brisbane, Australia

**Short title:** Validation of app to map social networks

**Keywords:** social network analysis, graph theory, Bluetooth, pervasive computing

[*] Corresponding author. Address: Black Dog Institute, Hospital Road, Randwick, NSW 2031, Australia; Phone: +61 2 9382 8507; Email: t.boonstra@unsw.edu.au

**Abstract**

Social network analysis is a prominent approach to investigate interpersonal relationships. Most studies use self-report data to quantify the connections between participants and construct social networks. In recent years smartphones have been used as an alternative to map networks by assessing the proximity between participants based on Bluetooth and GPS data. While most studies have handed out specially programmed smartphones to study participants, we developed an application for iOS and Android to collect Bluetooth data from participants' own smartphones. In this study, we compared the networks estimated with the smartphone app to those obtained from sociometric badges and self-report data. Participants (n=21) installed the app on their phone and wore a sociometric badge during office hours. Proximity data was collected for 4 weeks. A contingency table revealed a significant association between proximity data ($\phi$ = 0.17, p<0.0001), but the marginal odds were higher for the app (8.6%) than for the badges (1.3%), indicating that dyads were more often detected by the app. We then compared the networks that were estimated using the proximity and self-report data. All three networks were significantly correlated, although the correlation with self-reported data was lower for the app ($\rho$ = 0.25) than for badges ($\rho$ = 0.67). The scanning rates of the app varied considerably between devices and was lower on iOS than on Android. The association between the app and the badges increased when the network was estimated between participants whose app recorded more regularly. These findings suggest that the accuracy of proximity networks can be further improved by reducing missing data and restricting the interpersonal distance at which interactions are detected.

# 1. Introduction

Social network analysis is widely used to quantify relationships between people. Traditionally, social networks are mapped using survey data by simply asking respondents to identify their friends (e.g. [1]). These data are then used to define the edges between nodes in the network. This approach is time consuming, relies on subjective data collection, and is sensitive to the precise framing of the questions. New technologies have the potential to collect vast amounts of objective data at low cost and enable ecological momentary assessment, i.e. monitoring and assessment in real-time and real-world conditions [2]. Badges with embedded sensors have been used to record objective data on face-to-face interaction and proximity networks [3, 4]. Similarly, sensor-enabled smartphones can be used to map social networks by assessing physical proximity using either Bluetooth [5], location data [6] or by combining different data modalities [7-11]. While most studies have handed out specially programmed smartphones to study participants, we developed a smartphone application that participants can install on their own phone [12]. Using people's own smartphones may help scale up this technology for large-scale and population applications in research studies or self-monitoring.

The use of sensor technology to efficiently map dynamic social interactions has been well established. However, the large variety of different methods available also raises the need to cross-validate findings across technologies, types of devices and social settings (see also [13]). It has been shown that proximity data can be used to accurately infer friendships between participants [5]. However, several factors may affect both the networks that are derived from sensor data as well as the friendship reports obtained using surveys. For example, it has been shown that the size and characteristics of social networks vary considerably depending on the formulation of the questions use as name generators [14-16]. Similarly, when defining networks based on the frequency of email exchange, different choices of the threshold correspond to dramatically different network structures [17]. Similarly, missing data in survey collection [18] or passive smartphone data collection [12] can significantly impact on the resulting networks. Missing data is expected when sensor data is collected from smartphones in real life, where changes in data connectivity are frequent and participants turn off their device to preserve battery [7].

We developed an application for iOS and Android to passively collect Bluetooth data and map social networks of proximity. Here we assess the validity of social networks that are estimated based on Bluetooth data acquired using people's own smartphones. We cross-validated these against networks obtained using sociometric badges and self-reported survey data. The study aims to identify potential strengths and limitations of the technology that can inform larger studies on the role of social networks in mental health. The ability to accurately map social networks of proximity on a range of different smartphone types – across both Android and iOS operating systems – would enable using these technologies at scale.

## 2. Materials and Methods

2.1. Participants

Staff and research students within the Black Dog Institute in Sydney, Australia, were invited to join the study via an email sent to the general distribution list. The email contained a link to a participant information sheet. If they were interested in participating, they received a link to install the app on their smartphone. When the app was opened for the first time, participants were asked to complete a consent form included in the app. This study was approved by the University of New South Wales Human Research Ethics Committee (HC15202).

2.2. Procedure

The installed app included a short survey at the beginning of the study and then passively collected Bluetooth data for a four-week period (17 August to 11 September 2015) – see below for more details. Participants were also asked to wear a sociometric badge during office hours when present at the institute. Sociometric badges are wearable electronic badges that automatically measure the amount of face-to-face interaction, conversational time, physical proximity to other people, and physical activity levels to capture individual and collective patterns of behaviour [19]**.** To test the smartphone app, we investigated the scanning statistics and compared the social network estimated from proximity data (Bluetooth) against the social network estimated using the sociometric badges. In addition, we compared both proximity-based networks with the social network obtained using a name generator included in the survey.

2.3. Data acquisition

Native applications were developed for the Android and iOS operating systems, based on the results of our initial feasibility study [12]. For the iOS application we used the *BluetoothManager* private API, as the public *CoreBluetooth* API only contains functions for interacting with low-energy devices and it is currently not feasible to use Bluetooth Low Energy (BLE) to map social networks in iOS [20]. Both Android and iOS versions of the application asked the user to give consent prior to the commencement of Bluetooth data collection.

The application was configured to perform a Bluetooth discovery scan every five minutes during the study period. Bluetooth is a short-range communication protocol designed to allow a wireless connection between nearby devices. A key feature of a Bluetooth device is the ability to scan for other nearby devices. When a Bluetooth device conducts a discovery scan, other Bluetooth devices within a range of 5–10 m respond with their user-defined name, the device type, and a unique 12-hexadecimal-digit hardware media access control (MAC) address. A device's MAC address is fixed and can be used to differentiate one device from another. When a participant's MAC address is discovered by a periodic Bluetooth scan performed by another participant, it indicates that the two smartphones are within 5–10 m of each other (see also [12]).

As the Bluetooth MAC address of a device is potentially personally identifiable information, these data were cryptographically hashed on the handset to ensure the privacy of participants. Hashing generates a consistent 'signature' for each data item that cannot be reversed to reveal the original data value. In order to recreate the network and to distinguish participants from non-participants, devices would need to report their own MAC address. Since iOS devices were not able to retrieve their own MAC address, a helper system was designed using BLE. On iOS devices, the BLE service broadcast the device name and a writable characteristic for the MAC address. If the iOS device was in range of an Android device, the Android device would find the associated MAC address from the periodic Bluetooth scan and send it back to the iOS device. At the end of the study, all devices were able to report their own MAC address, ensuring a complete network of participants could be constructed.

The app was configured to collect data only during standard office hours (Monday-Friday, 9am-5pm). For each participant, the period over which the app was 'active' was retrospectively calculated based on the period over which Bluetooth discovery scans were initiated, the device was discovered by other participants, or the app's internal telemetry was recorded (e.g. to monitor battery usage).

Participants were also asked to fill out a short survey on the smartphone app at the start of the study. The survey included basic demographic questions (age and gender) as well as the opportunity to generate the names of up to five colleagues with whom they spend the most time to perform their job requirements. These data were used to construct their self-reported social network.

In addition to the smartphone app, sociometric badges (Sociometric Solutions, Boston MA) were used to record proximity networks, also using Bluetooth [19]. The badges contain a 2.4-GHz wireless transceiver (Chipcon, CC2500) and a class 2.0 Bluetooth module (BlueRadios, BR-46AR) for the detection of other badges in close proximity. The badges can also be configured to record additional sensor data, e.g. line of sight proximity using infrared emitters and detectors, audio recordings of speech, or accelerometer data. The badge was considered 'active' on the period over which Bluetooth, audio or acceleromoter data were available, although only the Bluetooth data were stored for analysis.

2.4. Connectivity analysis

Although the app also detects other Bluetooth devices, we only analysed the connectivity between participants. We estimate the connectivity between participants based on the Bluetooth scanning statistics of their smartphones. From these statistics, we define the connection strengths between participants and thus the weights of the network. The average connection strength between device *i* and *j* can then be represented as

$$R_{ij} = \frac{N_{ij}(T) + N_{ji}(T)}{N_i(T) + N_j(T)},\qquad(1)$$

where $N_{ij}$ is the number of scans where device *i* detected device *j* and $N_i$ the number of times device *i* scanned on time interval $T$ (see also [12]). By normalising the number of times one of the devices detected the other by the number of times each device scanned, the connection strength $R_{ij}$ is bound on the interval [0,1], where 1 indicates that both devices always detected each other when they scanned and 0 indicates that the devices never detected each other. If both devices did not scan during the interval of interest, $R_{ij}$ is set to zero.

2.5. Statistical analysis

We first compared the proximity data obtained using the smartphone app and the sociometric badges. A contingency table was created by comparing the time points at which the app or the badges detected a dyad (connection between two participants). To this end we pooled the data across all dyads. The contingency table quantifies the likelihood that when a dyad is detected by the app it is also detected by the badges and vice versa. We then estimated the association between these two binary variables using the odds ratio and the phi coefficient. Statistical significance was assessed using the Chi-squared test.

After directly comparing the proximity data obtained using the app and badges, we then compared the social networks that we constructed from these data. The undirected weights of the network were estimated by quantifying the percentages of time of the study period a dyad was detected, i.e. the devices of both participants were in close proximity. We used the Mantel test to quantify the association between the weighted adjacency matrices obtained using the app and badges. The Mantel test quantifies the correlation between matrices and uses permutation test to quantify statistical significance [21]. We used Spearman correlation and open-source code to quantify statistical significance [22] and used bootstrapping to quantify the 95% confidence interval of the correlation coefficient [23].

We then used the Mantel test to quantify the correlation between the adjacency matrices from the app and badges with the adjacency matrix obtained from the survey data. The survey data generated directed binary networks, which we first converted to an undirected network by collapsing directed edges between two nodes into a single undirected edge. We then extracted a binary backbone network from the weighted networks obtained using the app and badges [24]. This filtering method provides a statistical method to extract the relevant connection backbone in complex multiscale networks by preserving edges that are statistically significant. We used the R package 'disparityfilter' to extract the backbone network [25]. Alpha was set such that the binary network has the same density as the network obtained from the survey data. The binary undirected networks were then compared again using the Mantel test.

## 3. Results

21 participants agreed to join the study; 9 participants used Android handsets and 12 participants used iPhones. We first examined the amount of time the app and the badges were active during office hours. On average the app was active for 79.1 ± 22.2 % (Android: 68.0 ± 24.5 %, iOS: 87.5 ± 17.0 %) of office hours. Twelve of the 21 smartphone apps were active for more the 90% of the time (Fig. 1). The badges were active for an average of 37.0 ± 18.4% of office hours. The percentage of active time of the badges were active are generally lower than for the app, possibly because participants were only asked to wear the badges when they were at the office, while the app would be active regardless of the location of the participants.

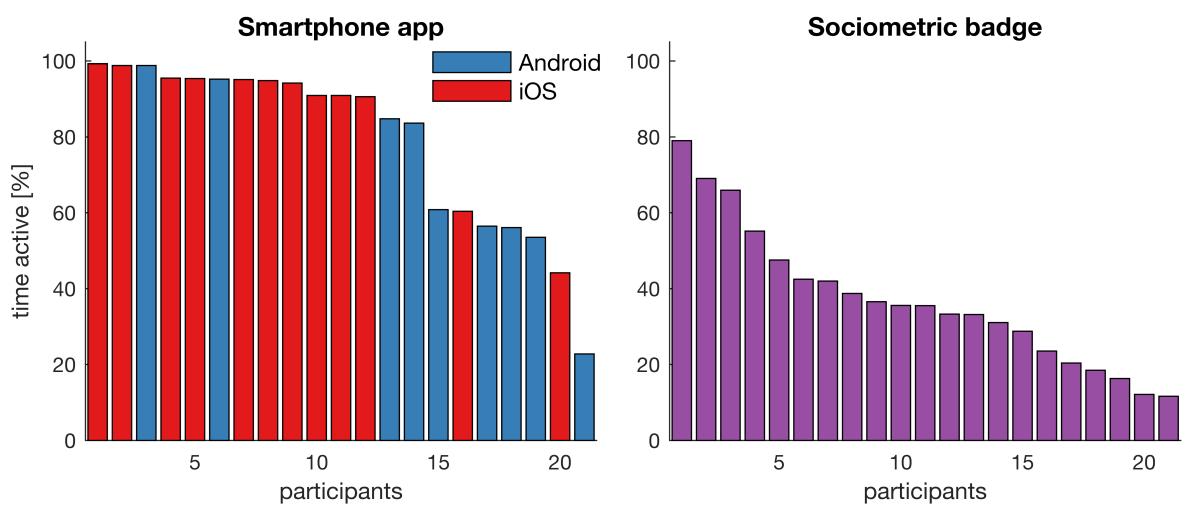

**Fig. 1. Percentages of study time the app and badges were active**. Left panel shows the activity of the smartphone app. Right panel shows the activity of the sociometric badges. Only office hours (Mon-Fri 9am-5pm) during the 4-week period were considered.

The scanning behaviour differed considerably between smartphones (Fig. 2A). Smartphones running Android scanned more often (on average 5.6 ± 3.8 scans per hour) than smartphones running iOS (1.1 ± 0.8 scans per hour). As proximity is an undirected measure, we can combine the data from smartphones A and B to estimate the edge between A and B. Of all 210 edges, 91.9% of the edges were on average scanned at least once each hour and 54.3% of the edges at least once every 15 minutes (Fig. 2B). Scanning rates of less than once every 15 minutes were mainly observed for connections between two iOS devices. The sociometric badges do not provide basic scanning statistics and we hence cannot determine how often the badges performed Bluetooth scans.

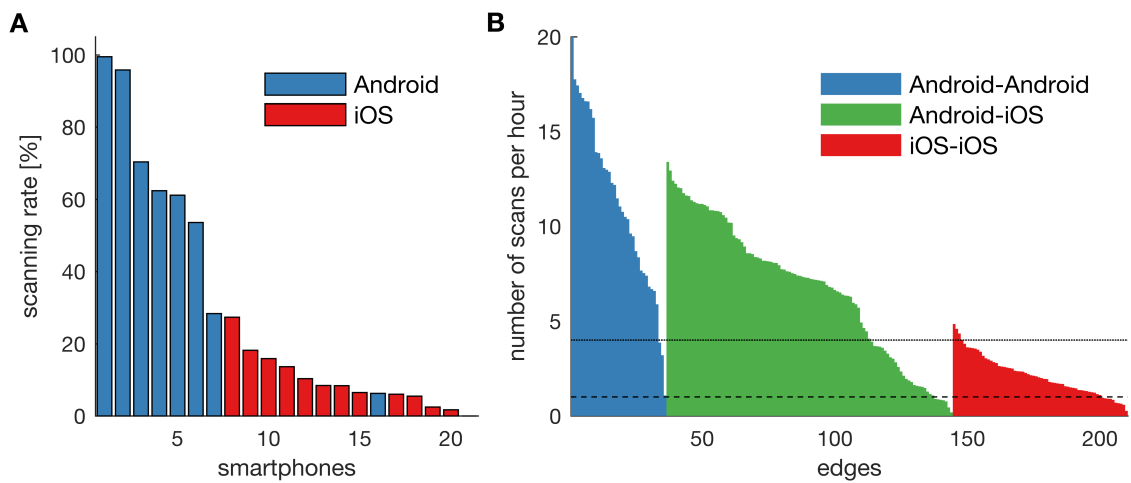

**Fig. 2. App scanning statistics**. **A)** Percentage of scheduled Bluetooth scans that were made by each smartphone in the 4-week period. **B)** Scanning rates for each edge of the network. The scanning rate between node A and B is determined by the number of scans made by smartphone A and B combined, as the edges are undirected (symmetric). The horizontal solid line reflects a scanning rate of 1 scan every 15 min; the dashed line 1 scan every hour.

To compare the smartphone app with the sociometric badges we first compared the time points at which the app or the badges detected a dyad. By pooling across all dyads, we constructed a contingency table of all time points at which a dyad was sampled by the app and then determine whether the badges detected the dyad at the same time point. Table 1 gives the contingency table across all office hours of the 4-week study period. The marginal odds show that physical proximity is sparse and that the app more often detected a dyad than the badges (2.92% for the app and 0.19% for the badges). Although the marginal odds differed considerable, there was a significant association between the time points at which the app and badges detected a dyad ($\phi = 0.10$, $\chi^2 =$ 2.5*10³, p<0.0001). The contingency table shows that the app more often detected a dyad when the badge did not (6448 times) than the other way around (264 times).

|     |      | Badge |        |
|-----|------|-------|--------|
|     |      | Hit   | Miss   |
| App | Hit  | 191   | 6448   |
|     | Miss | 264   | 227270 |

**Table 1. Contingency table across all office hours.** Table shows the number of times a particular edge of the network was detected (hit) or not (miss) by the smartphone app and the sociometric badges. Only office hours (Mon-Fri 9am-5pm) during the 4-week period were considered.

We then restricted the analysis to time intervals when both devices of a dyad were active. On average, the app on two smartphones was simultaneously active for 63.7 ± 26.0 % of the study time and pairs of sociometric badges were simultaneously active for 17.1 ± 11.7%. If we restrict the time interval to periods at which the app or badges of both participants were active, the marginal odds increased considerably (8.55% for the app and 1.28% for the badges). The association between both measures also increased ($\phi = 0.17$, $\chi^2 = 8.7*10^2$, p<0.0001). The app still more often detected a dyad when the badge did not (2327 times) than the other way around (214 times). We also compared the association separately for connections between Android users ($\phi = 0.11$, sensitivity = 0.69, specificity = 0.91), between iOS users ($\phi = 0.20$, sensitivity = 0.42, specificity = 0.94) and between Android and iOS users ($\phi = 0.18$, sensitivity = 0.49, specificity = 0.92).

|     |      | Badge |       |
|-----|------|-------|-------|
|     |      | Hit   | Miss  |
| App | Hit  | 191   | 2327  |
|     | Miss | 214   | 29252 |

**Table 2. Contingency table when the app and badge are both active.** Table shows the number of times a particular edge of the network was detected (hit) or not (miss) by the smartphone app and the sociometric badges. Only time intervals when both the app and badge were active were considered.

We then quantified the connection strength between participants by calculating the percentage of time participants were in close proximity. By estimating the connection strength between all pairs of participants the weighted adjacency matrix was obtained (Fig. 3A). We first estimated the adjacency matrix for the whole study duration (office hours during the 4-week period). The adjacency matrices again show that the app more often detected other devices than the sociometric badges (weighted network density: app = 0.024, badge = 0.002). The smartphone app detected some dyads for 45% of office hours, where the maximum connectivity for the badges was only 8% of the study time. Although they visually look quite different, the adjacency matrices of the app and badges were significantly correlated ($\rho$ = 0.25, 95% CI [0.24, 0.32], p = 0.0014). We then estimated the adjacency matrix only during time intervals when both devices of a dyad were active. As expected, only considering the time interval when both are active increased the percentage of time two devices detected each other (Fig. 3B). The correlation between the connectivity matrix of the app and badges remained largely the same ($\rho$ = 0.22, 95% CI [0.18, 0.29], p = 0.0038).

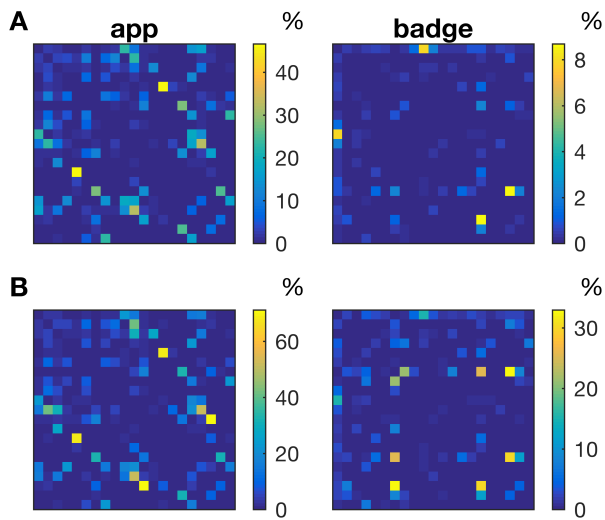

**Fig. 3. Weighted adjacency matrix of the social network mapped using the smartphone app and the sociometric badges.** Both axes reflect the 21 participants and each element reflects the percentage of time the two participants were in close proximity. **A)** Office hours, **B)** When both devices were active

We then compared the social networks of proximity with the networks derived from the survey data. The proximity networks are weighted undirected networks, whereas the survey data provide binary directed networks. To facilitate comparison, we converted the survey data into an undirected network and extracted a binary backbone network from the proximity data with the same network density. The three networks were all significantly correlated, although the correlation coefficient differed between pairs of networks (app-survey, ρ = 0.28, 95% CI [0.22, 0.35], p = 0.0005; badge-survey, ρ = 0.67, 95% CI [0.62, 0.72], p < 0.0001, app-badge, ρ = 0.28, 95% CI [0.23, 0.33], p = 0.0006). Counting the number of edges that matched between the networks, the network obtained using the app had fewer matching edges with networks from survey data (7/20) than the network obtained using the sociometric badges (14/20). Figure 4 shows the adjacency matrix and the topological representation of the three networks.

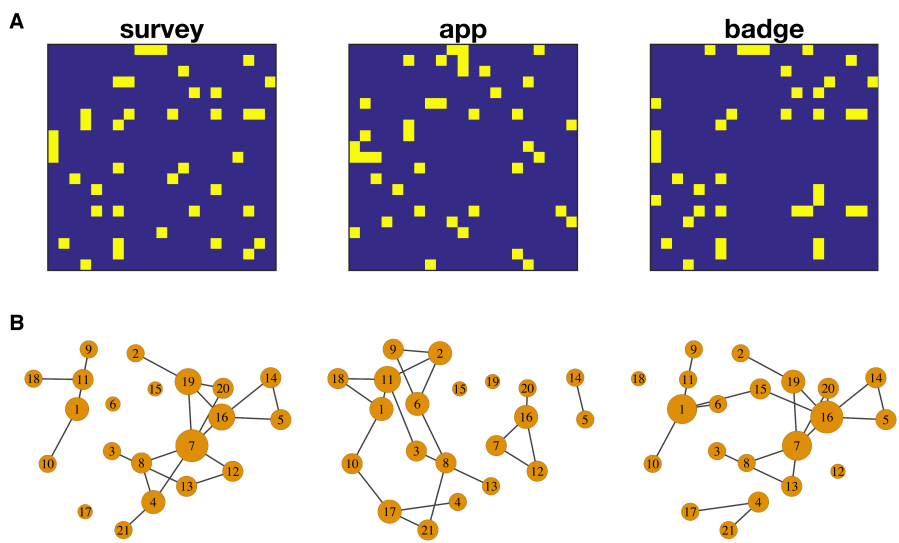

**Fig. 4. Social networks mapped using survey data, smartphone app and sociometric badges.** To facilitate comparison the incoming and outgoing edges of from the survey data were combined to obtain an undirected network. In addition, the binary backbone was extracted from the weighted adjacency matrix of the smartphone app and the sociometric badges. **A)** Adjacency matrices for the survey data, smartphone app and sociometric badges, **B)** Topological representation of the corresponding networks. Node size represents its degree. Layout was rendered using the Kamada-Kawai algorithm.

Finally, we used resampling to test for potential biases resulting from unequal scanning rates of the app. Weighted adjacency matrices were constructed by using a fixed number of random samples for each participant and correlated to the adjacency matrices constructed using the badge and survey data. Figure 5A shows the correlations coefficients for networks estimated with 10 to 500 random samples. As only the app data is resampled, the correlation between networks estimated using the badge and survey data remained fairly constant around 0.6. In contrast, the correlation between the networks estimated using the app and badge increased from 0.15 when only 10 samples were used to 0.51 when 500 samples were used. The correlation between the app and the survey fluctuated between 0.2 and 0.4. As only participants were included for which the app recorded the minimum number of required samples, the size of the networks decreased with increasing number of required samples (from 20 nodes at 10 samples to 10 nodes at 500 samples; Fig. 5B). As a result, the networks also become increasingly sparse: at 500 samples the survey network only contained 3 edges.

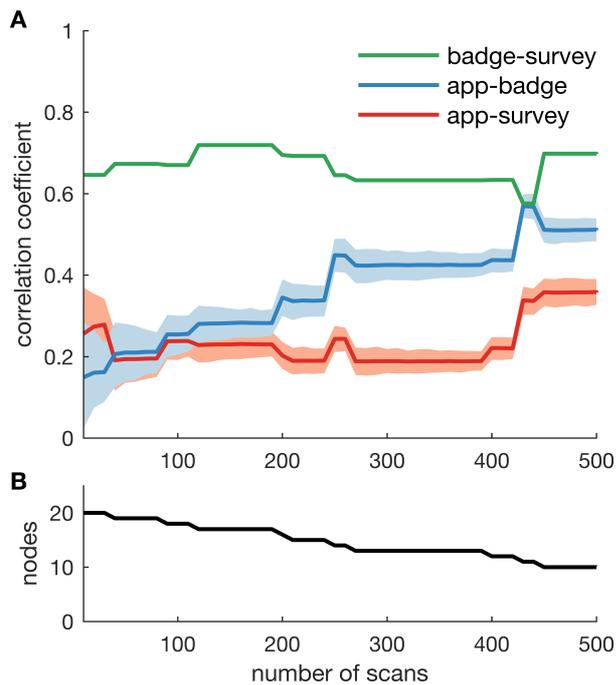

**Fig. 5. Correlation between resampled networks. A)** The weighted adjacency matrix of the app network was constructed using a fixed number of random samples for each participant to investigate potential biases resulting from unequal scanning rates. The number of required samples was varied from 10 to 500 samples. The Mantel test was again used to estimate the correlation with the networks constructed using badge and survey data. Colour patches show the 99% confidence interval estimated by resampling the network 1000 times. **B)** The size of the network that was compared decreased with increasing number of required samples, as participants with insufficient number of scans were excluded.

## 4. Discussion

We aimed to validate sensor technology to map social networks by comparing the proximity networks that were measured using a smartphone app and sociometric badges. The app and the badges both collected Bluetooth data and a name generator was used to map participants' self-reported social network. The app performed more frequent Bluetooth scans on Android devices (on average 5.6 scans per hour) than on iOS devices (1.1 scans per hour). The sociometric badges did not report basic scanning statistics but based on missing data points we determined that the badges were active for 37% of the study duration compared to 79% for the app. A contingency table revealed that the app was more like to detect a dyad than the badges: marginal odds 2.92% for the app and 0.19% for the badges. The weighted adjacency matrices obtained using the smartphone app and sociometric badges were significantly correlated ($\rho$ = 0.22-0.25). We then extracted the binary backbone networks from the weighted adjacency matrices to compare them with the self-reported networks. The binary network obtained using the badges was more strongly associated with the self-reported network ($\rho$ = 0.67) than the binary network obtained using the app ($\rho$ = 0.28). Although the association between social networks was statistically very robust, the proximity networks obtained using the smartphone app and the sociometric badges differed considerably. The association increased when the network was only estimated between participants whose app recorded at least 500 Bluetooth samples ($\rho$ = 0.51, n = 10).

In this study, we only analysed Bluetooth connectivity between devices from participants to enable the comparison between the smartphone app and the sociometric badges. It is interesting to note the different scanning statistics obtained using the badges and the app, although both are based on Bluetooth technology. Badges were only worn while in the workplace, whereas the app collected data during office hours regardless of location. Nevertheless, even when examining the data for the periods where both the badge and app were actively collecting data, the smartphone app provided a denser network than the badges. As the sociometric badges detected a dyad less often but revealed a stronger association with the self-reported networks, these findings may indicate that the Bluetooth range of the sociometric badges is smaller. That is, close proximity between participants may be a better proxy for actual social interactions and formal and informal interactions can be distinguished based on interpersonal distance [26]. As the Bluetooth range may be greater than the separation between rooms, it is also possible that dyads were detected between participants in neighbouring rooms and may therefore not just reflect face-to-face interactions between participants [19, 27]. As such, detected interactions at larger distances could be considered false positives, as they do not reflect true social interactions. Although the sociometric badges and most smartphones have class 2 Bluetooth with a range of about 10 m, differences in the Bluetooth radio hardware and software stacks used on different devices may result in different sensitivities and detection patterns [28]. The Received Signal Strength Indication (RSSI) of Bluetooth can be used to estimate distance between smart devices [29, 30]. However, RSSI is only available in the iOS *CoreBluetooth* API used for Bluetooth Low Energy (BLE) and not in *BluetoothManager* API used in the current study. It is currently not feasible to use BLE to map social networks, due to the inability of iOS devices to detect another iOS device when both are in a locked state [20].

Differences in network structure may also be partly due to participant behaviour, for example when someone carries their phone with them but leaves the badge behind in their office, or vice versa. The battery of the sociometric badges need to be regularly charged and badges need to be turned on when entering the office. Participants may forget to do this as they are less used to wearing and using the sociometric badges, which would result in missing data. The sociometric badges do not explicitly report missing Bluetooth data, but we considered the badges to be active on the period over which Bluetooth, audio or accelerometer data were available. This showed that the badges were only active for 37% of office hours, suggesting that participants regularly forgot to turn on or charge the sociometric badges. Although the app was active most of the time (79% of office hours), the scanning rates different considerable across devices and was much lower on iPhones (1.1 scans per hour on average) than smartphones running Android (5.6 scans per hour). The lower scanning rates on iPhones result from restrictions imposed by iOS on the background execution of apps, restricting the scope for passive data collection applications compared with Android devices.

Reduced scanning rates may affect the reliability of the estimated social network, in particular between devices that both have a reduced scanning rates (for example between two iOS devices). In a previous study, we showed that variations in scanning behaviour may introduce a bias in the estimation of social networks [12]. Although the scanning rates on iOS have improved compared to the previous study (1.1 vs. 0.35 scans per hour), this needs to increase further to improve the accuracy of the social network that can be mapped. Indeed, when we only used devices that recorded at least 500 Bluetooth samples the correlation with the networks estimated using the badges increased to 0.51. The current study has a small sample size (n=21), which reduces the precision of the correlation coefficients that are estimated. Due to the limited sample size, we cannot systematically test how the app performs on different types of smartphones and when running different versions of the operating system. Future studies involving larger samples can address individual variability and estimate the effect of user behaviour using subgroup analyses, for example to investigate potential gender differences in the estimation of proximity networks.

In the current study, we quantified the connection strength between participants as the percentage of time one of the devices is detected by another device. This simple metric may not be the best predictor of social connectivity and further feature engineering may assist in extracting the most important network features, for example by quantifying the duration or the frequency of contacts. The smartphone app and the sociometric badges collect dynamic connectivity data and the temporal patterns of social interactions provides valuable information about human social activity [11, 31]. Indeed, by using the temporal and spatial patterns of physical proximity data it is possible to accurately infer 95% of the self-reported friendships [5]. Several computational approaches have already been identified to improve the accuracy of information about social activities that can be derived from passively collected proximity data. For example, computational models have been used to identify both missing and spurious interactions and reconstruct a network that yields more accurate estimates of the true network properties than those provided by the observations themselves [32, 33]. In addition, probabilistic models can be used to discover interaction types from large-scale network data and infer the latent meaning of each interaction based on the set of observed interactions over slices of time [34]. These analytic tools may allow inference of the self-reported social connections more accurately from the proximity data than we collected in this study. However, self-reported social connections cannot be considered the gold standard, as this approach is subjective and depends on the type of name generators that are used [14-16]. Future research is hence needed to determine the relationship between different methods of mapping social networks.

## 5. Conclusion

The current findings show significant correlations between the social networks estimated using a smartphone app, sociometric badges and self-reported data, cross-validating these technologies to estimate proximity networks. Despite statistical robust correlations, large differences in networks were observed. These differences are most likely due to missing data, differences in range and participants not always carrying the devices with them. Sociometric badges were active for less than 40% of office hours, suggesting that participants often forgot to turn on or recharge their badges. In contrast, the smartphone app was active most of the time, but revealed a high rate of missing data in particular on iOS. Although background execution of apps is challenging on iOS, this is a technical problem that can likely be resolved through software engineering. Estimating proximity or restricting the range at which devices detect each other may enable more accurate information about social interactions, but the sociometric badges and the smartphone app do not have this functionality. BLE would allow to estimate the distance between devices, but this is currently not feasible on iOS. User behaviour is more difficult to control and some missing or spurious data is unavoidable, as participants will not always carry the device with them or forget to charge them. Recording over longer intervals and the use of computational models may enable the detection of these behaviours. A smartphone app is more convenient and less intrusive than devices build for research purposes. There are currently 2.3 billion smartphone users worldwide and is continuing to increase [35]. This study demonstrates that it is feasible to collect Bluetooth data on participants' own smartphones, rather than distributing devices to participants for the duration of the study. This has important implications on the ability to use this technology at scale, which is, for example, needed to reliably identify social markers of mental health [36, 37].

## Acknowledgments

The research was financially supported by NHMRC Centre of Research Excellence in Suicide Prevention APP1042580 and NHMRC John Cade Fellowship APP1056964.

**References**

1. Fowler JH, Christakis NA. Dynamic spread of happiness in a large social network: longitudinal analysis over 20 years in the Framingham Heart Study. Bmj. 2008;337:a2338. doi: 10.1136/bmj.a2338. PubMed PMID: 19056788; PubMed Central PMCID: PMC2600606.

2. Proudfoot J. The future is in our hands: the role of mobile phones in the prevention and management of mental disorders. The Australian and New Zealand journal of psychiatry. 2013;47(2):111-3. doi: 10.1177/0004867412471441. PubMed PMID: 23382507.

3. Wu L, Waber BN, Aral S, Brynjolfsson E, Pentland A. Mining face-to-face interaction networks using sociometric badges: Predicting productivity in an it configuration task. Available at SSRN 1130251. 2008.

4. Cattuto C, Van den Broeck W, Barrat A, Colizza V, Pinton J-F, Vespignani A. Dynamics of person-to-person interactions from distributed RFID sensor networks. PloS one. 2010;5(7):e11596.

5. Eagle N, Pentland AS, Lazer D. Inferring friendship network structure by using mobile phone data. Proceedings of the National Academy of Sciences of the United States of America. 2009;106(36):15274-8. doi: 10.1073/pnas.0900282106. PubMed PMID: 19706491; PubMed Central PMCID: PMC2741241.

6. Cho E, Myers SA, Leskovec J, editors. Friendship and mobility: user movement in location-based social networks. Proceedings of the 17th ACM SIGKDD international conference on Knowledge discovery and data mining; 2011: ACM.

7. Stopczynski A, Sekara V, Sapiezynski P, Cuttone A, Madsen MM, Larsen JE, et al. Measuring large-scale social networks with high resolution. PloS one. 2014;9(4):e95978. doi: 10.1371/journal.pone.0095978. PubMed PMID: 24770359; PubMed Central PMCID: PMC4000208.

8. Chronis I, Madan A, Pentland AS, editors. Socialcircuits: the art of using mobile phones for modeling personal interactions. Proceedings of the ICMI-MLMI'09 Workshop on Multimodal Sensor-Based Systems and Mobile Phones for Social Computing; 2009: ACM.

9. Miluzzo E, Lane ND, Fodor K, Peterson R, Lu H, Musolesi M, et al. Sensing meets mobile social networks: The design, implementation and evaluation of the CenceMe application. SenSys'08. 2008:337-50. PubMed PMID: WOS:000267822600025.

10. Jo H-H, Karsai M, Karikoski J, Kaski K. Spatiotemporal correlations of handset-based service usages. EPJ Data Science. 2012;1(1):1-18.

11. Sekara V, Stopczynski A, Lehmann S. Fundamental structures of dynamic social networks. Proceedings of the national academy of sciences. 2016;113(36):9977-82.

12. Boonstra TW, Larsen ME, Christensen H. Mapping dynamic social networks in real life using participants' own smartphones. Heliyon. 2015;1(3):e00037.

13. Pachucki MC, Ozer EJ, Barrat A, Cattuto C. Mental health and social networks in early adolescence: a dynamic study of objectively-measured social interaction behaviors. Social science & medicine. 2015;125:40-50. doi: 10.1016/j.socscimed.2014.04.015. PubMed PMID: 24797692.

14. Campbell KE, Lee BA. Name generators in surveys of personal networks. Soc Networks. 1991;13(3):203-21. doi: Doi 10.1016/0378-8733(91)90006-F. PubMed PMID: WOS:A1991GW01000001.

15. Eagle DE, Proeschold-Bell RJ. Methodological considerations in the use of name generators and interpreters. Soc Networks. 2015;40:75-83. doi: Doi 10.1016/J.Socnet.2014.07.005. PubMed PMID: WOS:000347585500008.

16. Marin A. Are respondents more likely to list alters with certain characteristics? Implications for name generator data. Soc Networks. 2004;26(4):289-307. doi: Doi 10.1016/J.Socnet.2004.06.001. PubMed PMID: WOS:000225137700001.

17. De Choudhury M, Mason WA, Hofman JM, Watts DJ, editors. Inferring relevant social networks from interpersonal communication. Proceedings of the 19th international conference on World wide web; 2010: ACM.

18. Kossinets G. Effects of missing data in social networks. Social networks. 2006;28(3):247-68.

19. Olguín DO, Waber BN, Kim T, Mohan A, Ara K, Pentland A. Sensible organizations: Technology and methodology for automatically measuring organizational behavior. IEEE Transactions on Systems, Man, and Cybernetics, Part B (Cybernetics). 2009;39(1):43-55.

20. Townsend S, Larsen ME, Boonstra TW, Christensen H. Using Bluetooth Low Energy in smartphones to map social networks. arXiv:150803938. 2015.

21. Mantel N. The detection of disease clustering and a generalized regression approach. Cancer research. 1967;27(2 Part 1):209-20.

22. Glerean E. Mantel test - Matlab implementation. Figshare. https://dx.doi.org/10.6084/m9.figshare.1008724.v3. Retrieved: 03 12, Nov 01, 2016. Figshare; 2014.

23. Efron B, Tibshirani R. Bootstrap methods for standard errors, confidence intervals, and other measures of statistical accuracy. Statistical science. 1986:54-75.

24. Serrano MÁ, Boguná M, Vespignani A. Extracting the multiscale backbone of complex weighted networks. Proceedings of the national academy of sciences. 2009;106(16):6483-8.

25. Bessi A, Briatte F. disparityfilter: Disparity Filter Algorithm for Weighted Networks. R package version 223. 2016;https://CRAN.R-project.org/package=disparityfilter.

26. Matic A, Osmani V, Mayora-Ibarra O. Analysis of social interactions through mobile phones. Mobile Networks and Applications. 2012;17(6):808-19.

27. Do TMT, Gatica-Perez D. Human interaction discovery in smartphone proximity networks. Personal and Ubiquitous Computing. 2013;17(3):413-31.

28. Semiconductor N. BLE on Android v1.0.1. . Retrieved from https://devzonenordicsemicom/attachment/bdd561ff56924e10ea78057b91c5c642. 2016.

29. Jung J, Kang D, Bae C. Distance estimation of smart device using bluetooth. Personal Computing Platform Research Team. 2013:13-8.

30. Rose A, Del Arroyo JG, Bindewald J, Ramsey B, editors. BlueFinder: A Range-Finding Tool for Bluetooth Classic and Low Energy. 12th International Conference on Cyber Warfare and Security 2017 Proceedings; 2017.

31. Holme P. Modern temporal network theory: a colloquium. The European Physical Journal B. 2015;88(9):1-30.

32. Guimerà R, Sales-Pardo M. Missing and spurious interactions and the reconstruction of complex networks. Proceedings of the National Academy of Sciences. 2009;106(52):22073-8.

33. Pan L, Zhou T, Lü L, Hu C-K. Predicting missing links and identifying spurious links via likelihood analysis. Scientific reports. 2016;6.

34. Do TMT, Gatica-Perez D, editors. Groupus: Smartphone proximity data and human interaction type mining. Wearable Computers (ISWC), 2011 15th Annual International Symposium on; 2011: IEEE.

35. Statista. https://www.statista.com/statistics/330695/number-of-smartphone-users-worldwide/ 2017.

36. Mohr DC, Zhang M, Schueller S. Personal sensing: Understanding mental health using ubiquitous sensors and machine learning. Annual Review of Clinical Psychology. 2017;13(1).

37. Boonstra TW, Werner-Seidler A, O'Dea B, Larsen ME, Christensen H. Smartphone app to investigate the relationship between social connectivity and mental health. Proceedings of the 39th Annual International Conference of the IEEE Engineering in Medicine and Biology Society. 2017;in press.